\documentclass[a4paper,12pt]{article}

\usepackage[left=2.5cm,right=2.5cm,top=2.5cm,bottom=2.5cm]{geometry}
\usepackage{graphicx,amssymb,amsmath,amsthm,setspace,subcaption,cite,authblk}

\usepackage[colorlinks=true,bookmarks=true]{hyperref}
\usepackage{breakurl} 

\usepackage{mathtools}

\DeclarePairedDelimiterX\braket[2]{\langle}{\rangle}{#1 \delimsize\vert #2}

\hypersetup{citecolor=blue}
\newcommand{\dif}{\mathrm{d}}
\newcommand{\Eqref}[1]{(\ref{#1})}
\newcommand{\half}{\frac{1}{2}}

\newcommand{\brac}[1]{\left(#1 \right)}
\newcommand{\sbrac}[1]{\left[#1\right]}
\newcommand{\im}{\mathrm{i}}
 
\begin{document}

\title{Motion of charged particles around a magnetic black hole/topological star with a compact extra dimension}

\author{Yen-Kheng Lim\footnote{Email: yenkheng.lim@xmu.edu.my}}

\affil{\normalsize{\textit{Department of Physics, Xiamen University Malaysia, 43900 Sepang, Malaysia}}}

\date{\normalsize{\today}}
\maketitle
 
\renewcommand\Authands{ and }
\begin{abstract}
 We study the motion of charged particles in a family of five-dimensional solutions describing either a black hole or topological star with a fifth compact dimension stabilised by a magnetic flux. The particle's trajectory is shown to move along the surface of a Poincar\'{e} cone. The radial motion shows a rich structure where the existence of various bound, plunging, or escaping trajectories depend on the constants of motion. Curves of energy and angular momentum corresponding to spherical orbits show a swallow-tail structure highly reminiscent to phase transitions of thermodynamics. When the momentum along the compact direction is varied, the is a critical point beyond which the swallow-tail kink disappears and becomes a smooth curve. 
\end{abstract}

\section{Introduction} \label{intro}

The study of geodesics and particle motion within a spacetime is an important tool to understand the properties of a gravitational system. For instance, the stability of circular orbits is, in specific cases, linked to quasinormal modes of the spacetime \cite{Cardoso:2008bp}.\footnote{However, see Ref.~\cite{Konoplya:2017wot} about the subtleties regarding the precise relation between the two.} From an astrophysical standpoint, the size of innermost stable circular orbits of charged particles around a black hole may reveal features of magnetic fields in its vicinity \cite{Frolov:2014zia}. In fact, as the direct imaging of a black hole is now a reality \cite{Akiyama:2019cqa}, one can seriously consider the optical appearance of the black hole, which can be calculated by considering null geodesics around it \cite{Luminet:1979nyg,Johannsen:2015hib,Luminet:2018wau,Wei:2019pjf}. Through this way, models of gravity can be tested or constrained. For instance, this has been done for braneworld gravity \cite{Vagnozzi:2019apd} and gravity coupled to non-linear electrodynamics \cite{Allahyari:2019jqz}.

In recent decades, theories of gravity with extra dimensions have received attention due to developments in theoretical physics such as string theory, holographic correspondences, and braneworld scenarios, among many others. A simple and natural candidate of a gravitating source in higher dimensional gravity would be a black hole with a compact fifth dimension. However, these black holes, including electrically charged ones, are known to suffer from Gregory--Laflamme instability \cite{Gregory:1993vy,Gregory:1994bj}. (See, e.g., Refs.~\cite{Kol:2004ww,Harmark:2005pp,Gregory:2011kh} for reviews.) 

The possibility of stable black holes with compact extra dimensions comes through a topological argument by Stotyn and Mann in \cite{Stotyn:2011tv}, where they argued that the presence of a magnetic charge may stabilise the spacetime.\footnote{On the other hand, one can also have stable black strings or more generally black $p$-branes, by considering Anti-de Sitter black strings or $p$-branes supported by scalar fields \cite{Cisterna:2017qrb}. These configurations were recently shown to be perturbatively stable \cite{Cisterna:2019scr}.} They introduced a solution to the five-dimensional Einstein--Maxwell equations characterised by two parameters, which are denoted $(\alpha,\beta)$ in the notation of the present paper. If $\alpha>\beta$, the spacetime carries a horizon and hence describes the black hole with the compact extra dimension stabilised by the magnetic charge. 

On the other hand, if $\alpha<\beta$, it describes a type of soliton star \cite{Stotyn:2011tv}. In this case, if a certain minimum radial distance is approached the spacetime caps off in a `cigar-like' geometry. Conical singularities may be present, unless the periodicity of the compact fifth dimension is appropriately fixed. In Refs.~\cite{Bah:2020ogh,Bah:2020pdz}, Bah and Heidmann allow the presence of orbifold fixed points. This introduces topological cycles in the compact fifth direction, and these solutions were called \emph{topological stars} by the authors. In this case, the magnetic field determines the minimum radial distance where the spacetime caps off. When the magnetic field is zero, this does not happen as the spacetime is simply a direct product between a Schwarzschild/Minkowski spacetime and a circle.

In this paper, we study the motion of charged particles in this family of solutions described in the preceding paragraphs. We shall use the terminology \emph{magnetic Kaluza--Klein black hole} (KKBH) to refer to the case $\alpha>\beta$, and \emph{magnetic topological star} (TS) to refer to the case $\alpha<\beta$. Since we always consider a non-zero magnetic field throughout the paper, we will often drop the term `magnetic' since it will be understood that it will be present at all times.

A guiding intuition in understanding the physics of this problem is the fact that the particle is under the influence of two forces. First is the spherically symmetric gravitational attraction towards the KKBH/TS, and second is the Lorentz force due to a spherically symmetric magnetic field. A similar situation occurs for a charged particle around a magnetically charged Reissner--Nordstr\"{o}m black hole, which was studied by Grunau and Kagramanova in \cite{Grunau:2010gd}. There, the authors obtained exact analytical solutions in terms of Weierstra\ss{} functions. Further studies of particles in the Reissner--Nordstr\"{o}m spacetime were subsequently done by other authors in \cite{Gad:2010,Pugliese:2011py,Sharif2017}. A non-relativistic analogue of this problem is the dyon-dyon interaction studied by Schwinger et al. in \cite{Schwinger:1976fr}. (See also \cite{Sivardiere2000,Balian:2005joa}.) In these similar/analogue problems, the motion of the electric charge is known to move on the surface of a Poincar\'{e} cone \cite{Poincare1896}.
 
It will be shown in this paper that the same is true for charged particles in the non-compact part of the KKBH/TS spacetime. The main reason for this similarity across all the aforementioned problems is that the equations of motion in the angular coordinates are the same.  They depend only on the electromagnetic properties of the system which are spherically symmetric, and independent of the other parameters that distinguishes the different systems. Relative to a choice of coordinate axes, the opening angle and orientation of the Poincar\'{e} cone depend on the product between the particle charge and the magnetic field strength, as well as its angular momentum. 
 
The radial motion of the particle can be classified into categories depending on whether it has access to the horizon, may escape to infinity, or bound in a finite domain. These depend on the particle's energy, momenta, and whether a horizon is present in the solution. If a horizon is present (the KKBH case) there is at most one bound domain where the particle orbits the black hole indefinitely. When a particle's energy exceeds a certain threshold, no bound orbits exist; it could either fall into the horizon or escape to infinity. In this sense, the situation is similar to particles around most spherically symmetric \cite{Hackmann:2008tu,Grunau:2010gd} as well as rotating black holes \cite{Wilkins:1972rs}. On the other hand, if no horizon is present (the TS case) the particle may have two disconnected bound domains. Furthermore, a bound domain may still exist when the energy exceeds the aforementioned threshold.
 
 The rest of the paper is organised as follows. In Sec.~\ref{sec_EOM} we describe the spacetime given by \cite{Bah:2020ogh,Bah:2020pdz} and derive the equations of motion for a charged particle in it. The parameter space of conserved quantities of the particle, as well as its domains of motion are studied in Sec.~\ref{sec_ParamSpace}. In particular, we obtain the Poincar\'{e} cone in Sec.~\ref{subsec_angular} and study the domains of radial motion in Sec.~\ref{subsec_r}. Conclusions and closing remarks are given in Sec.~\ref{sec_conclusion}. In this paper, we use geometrical units where the speed of light equals unity and the convention for Lorentzian metric signature is $(-,+,+,+,+)$.

\section{Equations of motion} \label{sec_EOM}

The five-dimensional KKBH/TS spacetime is described by the metric \cite{Stotyn:2011tv,Bah:2020ogh,Bah:2020pdz}
\begin{subequations} \label{metric}
\begin{align}
 \dif s^2&=-U\dif t^2+V\dif w^2+\frac{\dif r^2}{UV}+r^2\dif\theta^2+r^2\sin^2\theta\dif\phi^2,\\
 U&=1-\frac{\alpha}{r},\quad V=1-\frac{\beta}{r},
\end{align}
\end{subequations}
where $\alpha$ and $\beta$ are constants and $w$ is the coordinate representing the compact fifth dimension. If $\alpha>\beta$, the solution describes a KKBH and has a horizon at $r=\alpha$. On the other hand, if $\alpha<\beta$, the solution describes the TS spacetime where the spacetime caps off at $r=\beta$.

In this latter case, the periodicity of $w$ can be appropriately fixed to remove conical singularities, as was done in \cite{Stotyn:2011tv}, or such that certain orbifold singularities are allowed, as was done in \cite{Bah:2020ogh,Bah:2020pdz}. Here, we need not choose a particular periodicity for $w$ and mainly focus on the motion of particles in the non-compact directions, namely $(r,\theta,\phi)$.

The gauge potential of this solution is given by
\begin{align}
 A=-g\cos\theta\dif\phi. \label{gauge}
\end{align}
The Maxwell tensor is then obtained by taking the exterior derivative, $F=\dif A$. For the potential \Eqref{gauge}, the Maxwell tensor describes a spherically symmetric inverse-square magnetic field, whose strength is parametrised by $g$. The metric \Eqref{metric} with the gauge potential \Eqref{gauge} satisfies the Einstein--Maxwell equations in five dimensions provided that $g^2=\frac{3}{2}\alpha\beta$.\footnote{We normalise our gauge field such that the Einstein--Maxwell action appears as $I\propto\int\dif^5x\sqrt{-g}\brac{R-F_{\mu\nu}F^{\mu\nu}}$.}

The motion of a test particle of charge per unit mass $e$ is described by a spacetime curve $x^\mu(\tau)$, where $\tau$ is an appropriate affine parameter. Here we shall take our choice of parametrisation for $\tau$ such that 
\begin{align}
 g_{\mu\nu}\dot{x}^\mu\dot{x}^\nu=-1, \label{normalisation}
\end{align}
where over-dots denote derivatives with respect to $\tau$. The motion is governed by the Lagrangian $\mathcal{L}(x,\dot{x})=\half g_{\mu\nu}\dot{x}^\mu\dot{x}^\nu+eA_\mu\dot{x}^\mu$. For the spacetime described by \Eqref{metric} and \Eqref{gauge}, the Lagrangian is explicitly
\begin{align}
 \mathcal{L}(x,\dot{x})&=\half\brac{-U\dot{t}^2+V\dot{w}^2+\frac{\dot{r}^2}{UV}+r^2\dot{\theta}^2+r^2\sin^2\theta\dot{\phi}^2}-eg\cos\theta\dot{\phi}.
\end{align}
Since the magnetic field strength and the particle charge always appear together in the equations of motion, it will be convenient to define $q=eg$.  

The conjugate momenta are obtained by $p_\mu=\frac{\partial\mathcal{L}}{\partial\dot{x}^\mu}$. Explicitly, they appear as follows:
\begin{subequations}
\begin{align}
 p_t&=\frac{\partial\mathcal{L}}{\partial\dot{t}}=-U\dot{t},\\
 p_w&=\frac{\partial\mathcal{L}}{\partial\dot{w}}=V\dot{w},\\
 p_\phi&=\frac{\partial\mathcal{L}}{\partial\dot{\phi}}=r^2\sin^2\theta\dot{\phi}-q\cos\theta,\\
 p_r&=\frac{\partial\mathcal{L}}{\partial\dot{r}}=\frac{\dot{r}}{UV},\\
 p_\theta&=\frac{\partial\mathcal{L}}{\partial\dot{\theta}}=r^2\dot{\theta}.
\end{align}
\end{subequations}
Since our spacetime has three Killing vectors $\partial_t$, $\partial_w$, and $\partial_\phi$, the momenta along these directions are constants of motion. We shall denote the constants by 
\begin{align}
 p_t=-E,\quad p_w=P,\quad p_\phi=L,
\end{align}
representing the energy, linear momenta along the $w$-direction, and the angular momentum of the particle, respectively. The evolution of $t$, $w$, and $\phi$ are determined by the first integrals 
\begin{align}
 \dot{t}=\frac{E}{U},\quad \dot{w}=\frac{P}{V},\quad\dot{\phi}=\frac{L+q\cos\theta}{r^2\sin^2\theta}.\label{firstint1}
\end{align}
The equations of motion for $r$ and $\theta$ can be obtained from the Euler--Lagrange equations, giving
\begin{align}
 \ddot{\theta}&=-\frac{2\dot{r}\dot{\theta}}{r}+\frac{\cos\theta\brac{L+q\cos\theta}^2}{r^4\sin^3\theta}+\frac{q\brac{L+q\cos\theta}}{r^4\sin\theta}, \label{thetaddot}\\
 \ddot{r}&=\half\brac{\frac{U'}{U}+\frac{V'}{V}}\dot{r}^2+rUV\dot{\theta}^2-\frac{VU'E^2}{2U}+\frac{UV'P^2}{2V}+\frac{UV\brac{L+q\cos\theta}^2}{r^3\sin^2\theta}, \label{rddot}
\end{align}
where primes denote derivatives with respect to $r$.  

We also note that Eq.~\Eqref{normalisation} can be regarded as another equation of first integrals. Using Eq.~\Eqref{firstint1} to express $\dot{t}$, $\dot{w}$, and $\dot{\phi}$ in terms of the constants of motion, we have 
\begin{align}
 \frac{\dot{r}^2}{UV}+r^2\dot{\theta}^2-\frac{E^2}{U}+\frac{P^2}{V}+\frac{\brac{L+q\cos\theta}^2}{r^2\sin^2\theta}=-1. \label{firstint2}
\end{align}
Equations.~\Eqref{thetaddot} and \Eqref{rddot} can be solved numerically while using Eq.~\Eqref{firstint2} as a consistency check. In this work, this is performed by implementing the fourth-order Runge--Kutta algorithm in C.

A deeper analytical insight can be found by considering the Hamilton--Jacobi equation $\mathcal{H}\brac{\frac{\partial S}{\partial x},x}+\frac{\partial S}{\partial\tau}=0$, where $\mathcal{H}(p,x)=\half g^{\mu\nu}\brac{p_\mu-eA_\mu}\brac{p_\nu-eA_\nu}$ is the Hamiltonian obtained from the Legendre transform of the Lagrangian. Explicitly, the Hamilton--Jacobi equation for our present context reads
\begin{align}
 \half\Bigg[-\frac{1}{U}\brac{\frac{\partial S}{\partial t}}^2&+\frac{1}{V}\brac{\frac{\partial S}{\partial w}}^2+UV\brac{\frac{\partial S}{\partial r}}^2+\frac{1}{r^2}\brac{\frac{\partial S}{\partial\theta}}^2\nonumber\\
      &+\frac{1}{r^2\sin^2\theta}\brac{\frac{\partial S}{\partial\phi}+q\cos\theta}^2\Bigg]+\frac{\partial S}{\partial\tau}=0.
\end{align}
For this system, the Hamilton--Jacobi equation is completely separable, giving the first integrals 
\begin{subequations}\label{EOMdot}
\begin{align}
 \dot{t}&=\frac{E}{U},\quad \dot{w}=\frac{P}{V},\quad \dot{\phi}=\frac{L+q\cos\theta}{r^2\sin^2\theta},\label{twphidot}\\
 r^2\dot{\theta}&=\pm\sqrt{Q+L^2-\frac{\brac{L+q\cos\theta}^2}{\sin^2\theta}},\label{thetadot}\\
 r^2\dot{r}&=\pm\sqrt{r^4\brac{VE^2-UP^2}-r^2UV\brac{r^2+L^2+Q}}, \label{rdot}
\end{align}
\end{subequations}
where $Q$ is the Carter-like \cite{Carter:1968rr} separation constant. 

We further simplify the equations by introducing a Mino-type parameter \cite{Mino:2003yg} defined by $\frac{\dif\tau}{\dif\lambda}=r^2$, and changing variables to $x=\cos\theta$. Then the equations of motion now become
\begin{subequations}\label{HJE}
\begin{align}
 \frac{\dif t}{\dif\lambda}&=\frac{r^2E}{U}, \label{dt}\\
 \frac{\dif w}{\dif\lambda}&=\frac{r^2P}{V}, \label{dw}\\
 \frac{\dif\phi}{\dif\lambda}&=\frac{L+qx}{1-x^2},\label{dphi}\\
 \frac{\dif x}{\dif\lambda}&=\mp\sqrt{X(x)}, \label{dx}\\
 \frac{\dif r}{\dif\lambda}&=\pm\sqrt{R(r)}, \label{dr}
\end{align}
\end{subequations}
where 
\begin{subequations}
\begin{align}
 X(x)&=Q^2(1-x^2)-\brac{L+qx}^2,\label{Xdef}\\
 R(r)&=r^4\brac{VE^2-UP^2}-r^2UV\brac{Q+L^2+ r^2}.\label{Rdef}
\end{align}
\end{subequations}

\section{Parameter and coordinate ranges} \label{sec_ParamSpace}

\subsection{Angular motion} \label{subsec_angular}

The polar motion with $x=\cos\theta$ is governed by $X(x)$ in Eq.~\Eqref{Xdef}. Since $\frac{\dif x}{\dif\lambda}$ must be real in Eq.~\Eqref{dx}, the particle is allowed to move in the domain where $X(x)\geq 0$. Clearly, we see that no such domain exist if $Q+L^2\leq 0$. Therefore the Carter-like constant $Q$ is restricted to $Q>-L^2$. When this is satisfied, $X(x)$ is non-negative in the domain $x_-\leq x\leq x_+$, 
\begin{align}
 x_-\leq x\leq x_+, \label{domain_x}
\end{align}
where 
\begin{align}
 x_\pm&=\frac{-qL\pm\sqrt{(Q+q^2)(Q+L^2)}}{Q+L^2+q^2}.
\end{align}
The domain \Eqref{domain_x} is non-empty if $(Q+q^2)(Q+L^2)\leq 0$. Since we already argued above that $Q>-L^2$, we then have $Q+q^2\geq 0$. 
\begin{align}
 (Q,L)\in\left\{Q+L^2>0\quad\mbox{ and }\quad Q+q^2\geq0 \right\}. \label{domain_QL}
\end{align}

Equation.~\Eqref{dx} can be integrated explicitly upon a choice of branch and initial conditions. Here, let us consider two specific choices, $x(0)=x_+$ and $x(0)=x_-$. In the former, we take the upper (negative) sign of Eq.~\Eqref{dx}, whereas in the latter we take the lower (positive) sign. These choices will give an increasing $\lambda$ as the particle evolves away from their respective initial conditions. The integration is then\footnote{Note that in the argument of the cosine function, $\lambda$ lies outside the square root.}
\begin{align}
 \int_{x_\pm}^x\frac{\dif x'}{\sqrt{X(x')}}&=\mp\int_0^\lambda\dif\lambda'\nonumber\\
 x(\lambda)&=\frac{x_++x_-}{2}\pm\frac{x_+-x_-}{2}\cos\brac{\sqrt{Q+L^2+q^2}\lambda} \nonumber\\
 x(\lambda)&=\frac{1}{Q+L^2+q^2}\sbrac{-qL\pm\sqrt{(Q+q^2)(Q+L^2)}\cos\brac{\sqrt{Q+L^2+q^2}\lambda}}. \label{soln_x}
\end{align}
We can also obtain an analytical solution for $\phi$ expressed as a function of $x$ by eliminating $\lambda$ from Eq.~\Eqref{dphi} and \Eqref{dx}, giving 
\begin{align}
 \frac{\dif\phi}{\dif x}&=\mp\frac{L+qx}{(1-x^2)\sqrt{X(x)}}.
\end{align}
The integral can be performed with the aid of partial fraction decomposition on the factor $1/(1-x^2)$. The result is 
\begin{align}
 \phi(x)&=\brac{\mathrm{sgn}(L-q)+\mathrm{sgn}(L+q)}\frac{\pi}{4}\nonumber\\
    &\quad\pm\half\Bigg\{\mathrm{sgn}(L-q)\arcsin\sbrac{\frac{(L-q)^2}{\sqrt{(Q+q^2)(Q+L^2)}}\brac{\frac{1}{1+x}-\frac{Q+L^2+q^2-qL}{(L-q)^2}}}\nonumber\\
    &\quad\quad-\mathrm{sgn}(L+q)\arcsin\sbrac{\frac{(L+q)^2}{\sqrt{(Q+q^2)(Q+L^2)}}  \brac{\frac{1}{1-x}-\frac{Q+L^2+q^2+qL}{(L+q)^2}}}\Bigg\}, \label{soln_phi}
\end{align}
where the `$\pm$' signs are in accordance to the choice of initial conditions of $x(\lambda)$ in Eq.~\Eqref{soln_x}. We have also defined the sign function as $\mathrm{sgn}(x)$ which returns $\pm1$ if $x\gtrless 0$ and returns $0$ if $x=0$.

It is worth noting that the angular equations of motion are independent of $\alpha$ and $\beta$, and are purely due to the Lorentz interaction between the charge and the spherically symmetric magnetic field. As alluded to in the Introduction, the equations of motion for $x=\cos\theta$ and $\phi$ are in fact identical to the equations of motion for a charged particle around a magnetically charged Reissner--Nordstr\"{o}m black hole \cite{Hackmann:2008tu,Grunau:2010gd,Pugliese:2011py,Sharif2017}, as well as the non-relativistic dyon-dyon interaction \cite{Schwinger:1976fr,Sivardiere2000} where the motion is confined to the Poincar\'{e} cone.

We will show that the Poincar\'{e} cone also exists in our context of the KKBH/TS spacetime as well. If we take our coordinates $(r,\theta,\phi)$ to define a naive Euclildean three-space with $\{\hat{e}_r,\hat{e}_\theta,\hat{e}_\phi\}$ as the ortho-normal basis in spherical coordinates, the vector 
\begin{align}
 \vec{J}=-q\,\hat{e}_r - r^2\sin\theta\,\dot{\phi}\,\hat{e}_\theta+r^2\dot{\theta}\,\hat{e}_\phi
\end{align}
is conserved throughout the motion, i.e., $\frac{\dif}{\dif\tau}\vec{J}=\vec{0}$. In the non-relativistic case, this quantity is the total angular momentum of the system \cite{Schwinger:1976fr,Sivardiere2000,Balian:2005joa}. This implies that the particle moves on the surface of a cone which subtends an angle $\chi$ such that 
\begin{align}
 \cos\frac{\chi}{2}=-\frac{\vec{J}}{\big|\vec{J}\big|}\cdot\hat{e}_r=\frac{q}{\sqrt{q^2+Q+L^2}}. \label{cone_chi}
\end{align}
If we further define a Cartesian coordinate system in this naive Euclidean space by 
\begin{align}
 x_1=r\sin\theta\cos\phi,\quad x_2=r\sin\theta\sin\phi,\quad x_3=r\cos\theta,
\end{align}
the angle $\psi$ of the cone's axis with the $x_3$-direction (the axis passing through the north and south poles) is given by 
\begin{align}
 \cos\psi=\frac{\vec{J}}{\big|\vec{J}\big|}\cdot\hat{e}_3=\frac{L}{\sqrt{q^2+Q+L^2}}, \label{cone_psi}
\end{align}
where Eq.~\Eqref{EOMdot} was used and $\hat{e}_3$ is the unit vector along $x_3$.

%
%

\subsection{Radial motion} \label{subsec_r}

The motion in the $r$-direction is governed by the function $R(r)$ defined in Eq.~\Eqref{Rdef}, which we will rewrite here as
\begin{align}
 R(r)=c_4r^4+c_3 r^3+c_2r^2+c_1r+c_0, \label{R_polynomial}
\end{align}
where
\begin{subequations}
\begin{align}
 c_4&=-\brac{1+P^2-E^2},\\
 c_3&=\alpha+\beta-\beta E^2+\alpha P^2,\\
 c_2&=-\brac{\alpha\beta+Q+L^2},\\
 c_1&=(\alpha+\beta)\brac{Q+L^2},\\
 c_0&=-\alpha\beta\brac{Q+L^2}.
\end{align}
\end{subequations}
From Eq.~\Eqref{dr}, the requirement that $\frac{\dif r}{\dif\lambda}$ be real means the particle can only access the domains of $r$ where $R(r)\geq0$. We identify these domains by studying the root structure of $R(r)$, which serves as possible boundaries of the domains. 

It will be convenient to define $K=L^2+Q$, as $L$ and $Q$ always appear in this combination in $R(r)$. To aid our discussion below, we introduce the following terminology for the possible domains of $r$ such that $R(r)\geq0$:
\begin{itemize}
 \item \emph{Plunging orbits}. A finite domain of $r$ which contains the horizon $r=\alpha$. Particles in this domain may fall into the horizon.
 \item \emph{Escaping orbits}. A (semi-)infinite domain of $r$, where $R(r)$ remains positive as $r\rightarrow\infty$. Particles in this domain can escape to infinity.
 \item \emph{Bound orbits}. A finite domain bounded by two roots of $R(r)$. Particles in this domain are in stable bound orbits, neither falling into the black hole or escaping to infinity.
\end{itemize}

Suppose we start with a case where $R(r)$ has four real roots. Varying $E$, $P$, and $K$ will generally vary the positions of each root. A pair of roots will coalesce into a degenerate root when $R(r)=R'(r)=0$. A particle located at this point will will solve the equations of motion for constant $r$, which we will call a \emph{circular orbit}.\footnote{The term `circular' used here has an interesting roundabout connotation. Usually, constant-$r$ solutions of spherically symmetric equations of motion reduces to a circle because the symmetry confines the motion to a plane containing the origin. When spherical symmetry is not present, constant-$r$ trajectories may lie on a sphere, and are typically called \emph{spherical orbits} \cite{Teo:2020sey}. At the same time we have shown in the previous subsection that the trajectories lie on a Poincar\'{e} cone. The intersection between a cone and a sphere whose apex and centres coincide is, again, a circle.} This condition is satisfied when the energy and angular momentum satisfies
\begin{subequations}\label{EL_sph}
\begin{align}
 E^2=E^2_{0}&=\frac{(r-\alpha)^2\sbrac{2(r-\beta)^2+r^2(2r-3\beta)P^2}}{r(r-\beta)^2(2r-3\alpha)},\label{E_sph}\\
 L^2+Q=K_{0}&=\frac{r^2\sbrac{\alpha(r-\beta)^2+r^2(\alpha-\beta)P^2}}{(r-\beta)^2(2r-3\alpha)}. \label{L_sph}
\end{align}
\end{subequations}
For a given $\alpha$, $\beta$, and $P$, plotting Eq.~\Eqref{EL_sph} as a parametric curve in $r$ on the $E^2$-$K$ plane will serve as a boundary separating domains for which $R(r)$ has various numbers of real or complex roots.

We can identify which pair among the four roots are degenerate by evaluating the second derivative
\begin{align}
 R''(r)\Big|_{\substack{E=E_0,\\L=L_0}}&=-\frac{2\sbrac{\alpha(r-\beta)^3(r-3\alpha)+r^2(\alpha-\beta)\brac{r^2-3(\alpha+\beta)r+6\alpha\beta}P^2}}{(r-\beta)^2(2r-3\alpha)}.
\end{align}
If the double root of $R(r)$ is a local minimum, this corresponds to an unstable circular orbit. On the other hand, if this double root is a local maximum, the circular orbit is stable as a small perturbation puts it in small oscillations about its original radius. The \emph{critical circular orbit} (CCO) is $r_{\mathrm{CCO}}$ such that $R''(r_{\mathrm{CCO}})=0$. This is the critical radius where a circular orbit changes from being unstable to stable, or vice versa. Indeed, if $P=\beta=0$, and letting $\alpha=2m$, we recover the innermost stable circular orbit (ISCO) of the Schwarzschild black hole, $r_{\mathrm{ISCO}}=6m$.\footnote{As we shall see in the following, the reason we use the terminiology `critical circular orbit' rather than `innermost' is there may be multiple points satisfying $R''(r)=0$, and hence `innermost' is no longer accurate.}

Looking at large $r$, we find that, asymptotically,
\begin{subequations}\label{asymptoticEK}
\begin{align}
 E_0^2&\sim 1+P^2-\half\brac{\alpha+(\alpha-\beta)P^2}\frac{1}{r}+\mathcal{O}\brac{\textstyle{\frac{1}{r^2}}},\\
 K_0&\sim\half\brac{\alpha+(\alpha-\beta)P^2}r+\frac{1}{4}(3\alpha+4\beta)(\alpha-\beta)P^2+\frac{3}{4}\alpha^2+\mathcal{O}\brac{\textstyle{\frac{1}{r}}}.
\end{align}
\end{subequations}

Next we will explore the $E^2$-$K$ space as $\alpha$ varies, starting from $\alpha>\beta$ (the KKBH case), and then for $\alpha<\beta$ (the TS case).

\subsubsection{\texorpdfstring{$\alpha>\beta$}{alpha>beta} (KKBH)}

\begin{figure}
 \begin{center}
  \includegraphics{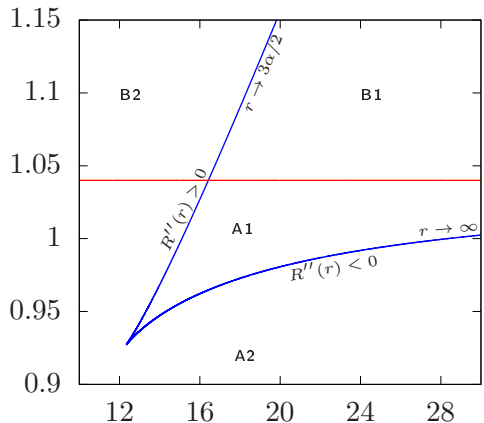}
  \caption{Parameter space described by angular momentum $L$ and energy $E$, plotted here for $\alpha=2$, $\beta=1$, and $P=0.2$. The blue curves corresponds to the circular orbits, where the second derivatives of $R(r)$ on these orbits are indicated. The segment with $R''(r)<0$ give stable circular orbits while and the other segment with $R''(r)>0$ give unstable circular orbits. The two segments meet at $r=r_{\mathrm{CCO}}$. The four domains marked A1, A2, B1, and B2 have functions $R(r)$ sketched in Figs.~\ref{fig_BH_domainA1}, \ref{fig_BH_domainA2}, \ref{fig_BH_domainB1}, and \ref{fig_BH_domainB2}, respectively.}
  \label{fig_BH_domain_LE}
 \end{center}

\end{figure}

We begin with $\alpha>\beta$, which correspond to spacetimes with an event horizon. The $E^2$-$K$ space can be organised into domains separated by curves $E^2=E_0^2$ and $K=K_0$, as defined in Eq.~\Eqref{EL_sph}. Fig.~\ref{fig_BH_domain_LE} demonstrates the domains for the concrete example $\alpha=2$, $\beta=1$, and $P=0.2$. 

For $\alpha>\beta$, we find that there are two branches of circular orbits. The unstable branch has $R''(r)>0$ and occurs for $\frac{3}{2}\alpha<r<r_{\mathrm{CCO}}$ and is the upper branch depicted in Fig.~\ref{fig_BH_domain_LE}. This branch tends to infinity as $r\rightarrow\frac{3}{2}\alpha$. On the other hand, the stable branch is the one with $R''<0$ and corresponds to $r_{\mathrm{CCO}}<r<\infty$. This branch asymptotically approaches $E^2=1+P^2$ as $r\rightarrow\infty$, in accordance to Eq.~\Eqref{asymptoticEK}.

The horizontal red line is $E^2=1+P^2$. Values of $E^2$ below this line makes the leading coefficient $c_4$ of $R(r)$ negative. We shall call this case `A'. Conversely, values of $E$ above this line shall be called case `B' for which $c_4$ is positive.

First we look at Case A. As the leading coefficient $c_4$ is negative,  we have $R(r)<0$ for values of $r$ beyond its largest root. This implies that the particle is unable to escape to infinity. The parameters giving four real roots of $R(r)$ lie in domain A1 in Fig.~\ref{fig_BH_domain_LE}. The structure of these four roots can be understood with the aid of the Descartes rule of signs. For $K=L^2+Q$ satisfying \Eqref{domain_QL}, along with $\alpha,\,\beta\geq0$, we see that $c_2<0$, $c_1>0$, and $c_0\leq 0$.

\begin{figure}
 \begin{center}
  \begin{subfigure}[b]{0.49\textwidth}
    \centering
    \includegraphics[scale=0.9]{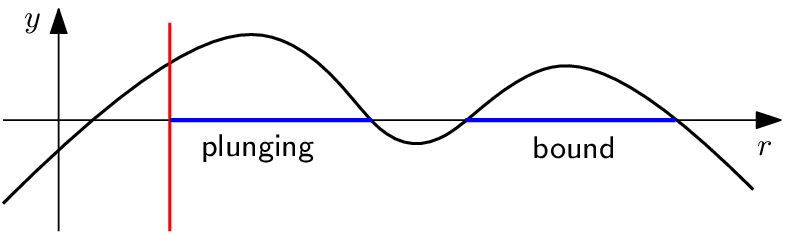}
    \caption{Case A1}
    \label{fig_BH_domainA1}
  \end{subfigure}
  \begin{subfigure}[b]{0.49\textwidth}
    \centering
    \includegraphics[scale=0.9]{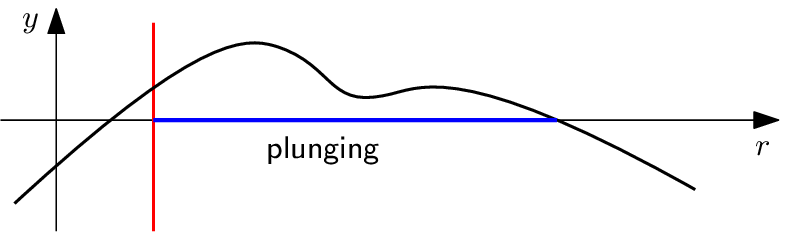}
    \caption{Case A2}
    \label{fig_BH_domainA2}
  \end{subfigure}
  \par
  \vspace{24pt}
  \begin{subfigure}[b]{0.49\textwidth}
    \centering
    \includegraphics[scale=0.9]{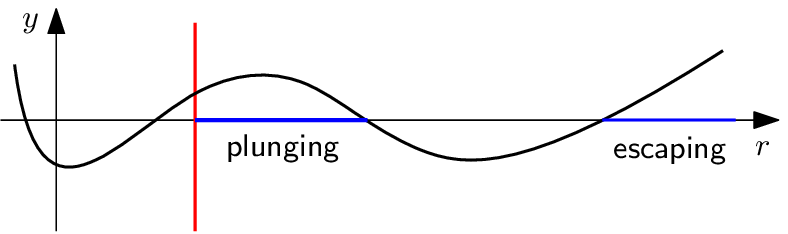}
    \caption{Case B1}
    \label{fig_BH_domainB1}
  \end{subfigure}
  \begin{subfigure}[b]{0.49\textwidth}
    \centering
    \includegraphics[scale=0.9]{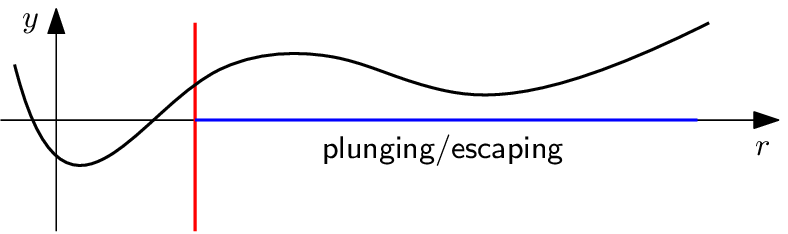}
    \caption{Case B2}
    \label{fig_BH_domainB2}
  \end{subfigure}
 
 \caption{Sketches of $y=R(r)$ for various cases of energy and angular momentum, corresponding to domains (a)--(d) of the $L$-$E$ plane of Fig.~\ref{fig_BH_domain_LE}. The vertical red lines indicate the horizon $r=\alpha$. The allowed domains of $r$ corresponding to $R(r)\geq0$ are marked as thick blue lines. For each allowed domain, the labels \emph{plunging}, \emph{escaping}, and \emph{bound} are as defined in the main text.}
 \label{fig_BH_domain}
 \end{center}

\end{figure}

The remaining coefficient $c_3$ can be rearranged as 
\begin{align*}
 c_3&=\beta\sbrac{\frac{\alpha}{\beta}\brac{1+P^2}-E^2}.
\end{align*}
Since $E^2<1+P^2$ and $\alpha>\beta$, this term is positive. By the Descartes rule of signs, $R(r)$ in this case has four positive roots. Since $R(\alpha)=E^2\alpha^3\brac{\alpha-\beta}>0$, the number of roots located outside the horizon is either 3 or 1. From these considerations, we conclude that, for $E^2<1+P^2$, there is always a plunging orbit and at most one bound orbit. A sketch of such a situation is shown in Fig.~\ref{fig_BH_domainA1}, which occurs when $E$ and $K$ takes values in the subdomain A1 of Fig.~\ref{fig_BH_domain_LE}. For $r>r_{\mathrm{CCO}}$, the two largest roots will coalesce, shrinking the domain of bound orbit to a point, giving a stable circular orbit. On the other hand, if $r<r_{\mathrm{CCO}}$, the second and third largest roots coalesce such that the plunging and bound domains are only separated by a point, which is the unstable circular orbit. For this case ($\alpha>\beta$), we see that it is indeed appropriate to call $r_{\mathrm{CCO}}$ the \emph{innermost} stable circular orbit.

Varying $E^2$ and $K$ further until the pair of roots become complex, the function $R(r)$ will appear as sketched in Fig.~\ref{fig_BH_domainA2}, where only a plunging orbit can exist. This occurs when $E^2$ and $K$ takes values in subdomain A2 of Fig.~\ref{fig_BH_domain_LE}.

We now turn to Case B, where $E^2>1+P^2$. Now $c_4$ is positive. Whether $c_3$ is positive or negative, the Descartes rule of signs tells us that $R(r)$ has either three positive roots or one positive root. Since $R(r)$ is positive at the horizon, we conclude that there are no bound orbits in this case. The possible types of motion are either plunging and/or escaping orbits. More specifically, there are two subdomains B1 and B2, where domain B1 corresponds to functions $R(r)$ appearing in the form sketched in Fig.~\ref{fig_BH_domainB1}, and domain B2 have functions appearing as sketched in Fig.~\ref{fig_BH_domainB2}.

Finally we look specifically at the circular orbits, which lie on the boundary curves separating the domains discussed above, as shown in Fig.~\ref{fig_BH_domain_LE}. In particular, the curve \Eqref{EL_sph} for $\frac{3}{2}\alpha<r<r_{\mathrm{CCO}}$ is the upper curve. Particles whose energy and angular momentum taking values along this upper curve are unstable spherical orbits. The lower curve gives stable circular orbits with radii in the range $r_{\mathrm{CCO}}<r<\infty$.

\subsubsection{\texorpdfstring{$\frac{2}{3}\beta<\alpha<\beta$}{2beta/3<alpha<beta} (TS)}

\begin{figure}
 \begin{center}
  \begin{subfigure}[b]{0.49\textwidth}
    \centering
    \includegraphics[scale=1]{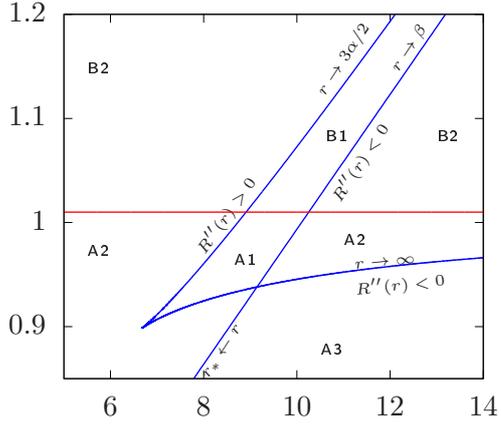}
    \caption{$P=0.1$.}
    \label{fig_top_ELspace_seq1}
  \end{subfigure}
  \begin{subfigure}[b]{0.49\textwidth}
    \centering
    \includegraphics[scale=1]{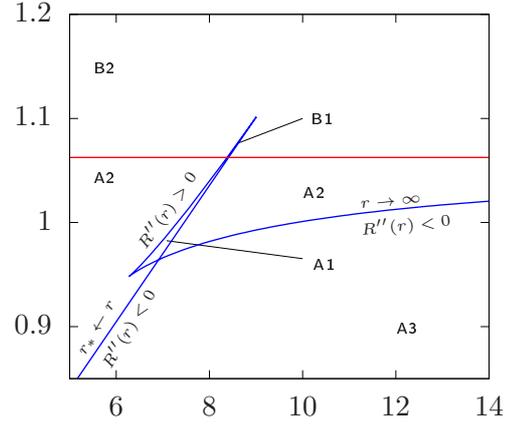}
    \caption{$P=0.25$.}
    \label{fig_top_ELspace_seq2}
  \end{subfigure}
  \begin{subfigure}[b]{0.49\textwidth}
    \centering
    \includegraphics[scale=1]{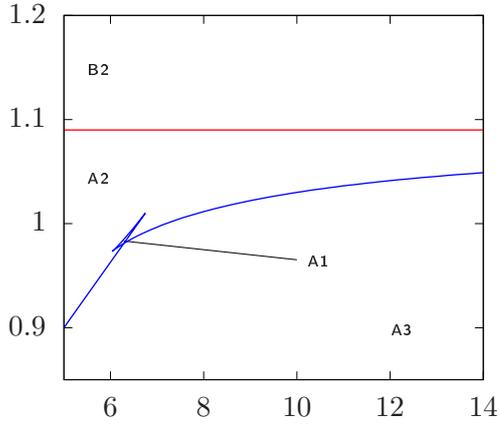}
    \caption{$P=0.3$.}
    \label{fig_top_ELspace_seq3}
  \end{subfigure}
  \begin{subfigure}[b]{0.49\textwidth}
    \centering
    \includegraphics[scale=1]{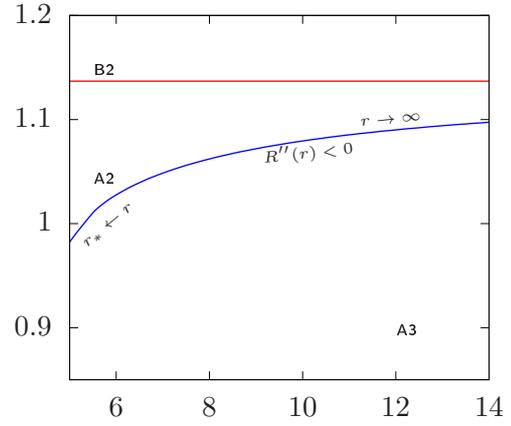}
    \caption{$P=0.36$.}
    \label{fig_top_ELspace_seq4}
  \end{subfigure}
 \caption{Parameter space for the TS spacetime with $\alpha=1.5$, $\beta=2$, and various $P$. The horizontal direction represents $L^2+Q$ and the vertical direction represents $E^2$. The horizontal red lines correspond to $E^2=1+P^2$.}
 \label{fig_top_ELspace_seq}
 \end{center}

\end{figure}

When $\beta>\alpha$, the spacetime describes a TS without the presence of a horizon. The structure of the parameter space in this case is richer, as there can be up to \emph{three} branches of circular orbits, depending on the value of $P$. Figure.~\ref{fig_top_ELspace_seq} shows the sequence of the curves \Eqref{EL_sph} for the concrete example $\alpha=1.5$, $\beta=2$, and increasing values of $P$.

\begin{figure}
 \begin{center}
  \begin{subfigure}[b]{0.49\textwidth}
    \centering
    \includegraphics[scale=0.9]{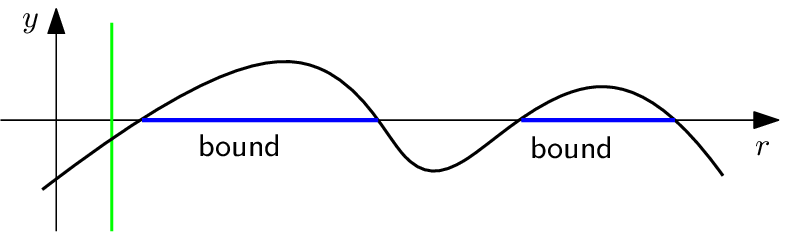}
    \caption{Case A1}
    \label{fig_top_domainA1}
  \end{subfigure}
  \begin{subfigure}[b]{0.49\textwidth}
    \centering
    \includegraphics[scale=0.9]{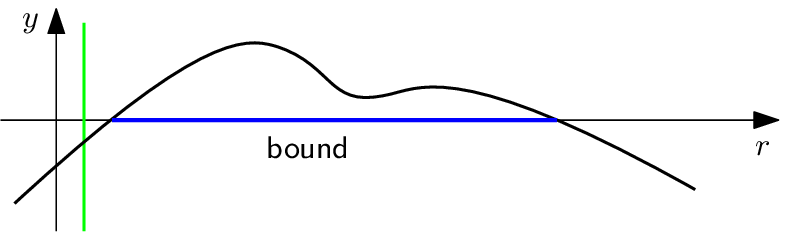}
    \caption{Case A2}
    \label{fig_top_domainA2}
  \end{subfigure}
  \begin{subfigure}[b]{0.49\textwidth}
    \centering
    \includegraphics[scale=0.9]{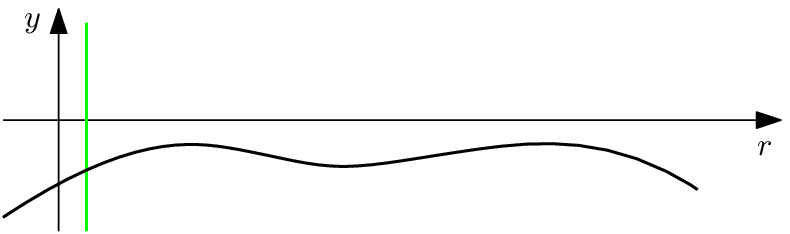}
    \caption{Case A3}
    \label{fig_top_domainA3}
  \end{subfigure}
  \\\vspace{12pt}
  \begin{subfigure}[b]{0.49\textwidth}
    \centering
    \includegraphics[scale=0.9]{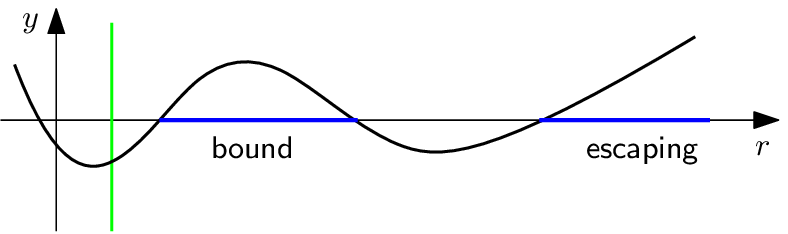}
    \caption{Case B1}
    \label{fig_top_domainB1}
  \end{subfigure}
  \begin{subfigure}[b]{0.49\textwidth}
    \centering
    \includegraphics[scale=0.9]{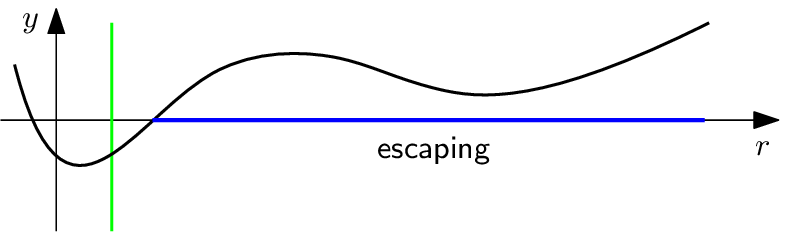}
    \caption{Case B2}
    \label{fig_top_domainB2}
  \end{subfigure}
 \caption{Sketches of $y=R(r)$ for various cases of energy and angular momentum, corresponding to cases A1--A3, B1--B2 of the $K$-$E^2$ plane of Fig.~\ref{fig_top_ELspace_seq}. The vertical green lines indicate $r=\beta$. The allowed domains of $r$ corresponding to $R(r)\geq0$ are marked as thick blue lines. For each allowed domain, the labels \emph{escaping} and \emph{bound} are as defined in the main text. Since there are no horizons in this case, there are no plunging orbits.}
 \label{fig_top_domain}
 \end{center}

\end{figure}

We shall attempt to understand these structures by looking closely at the circular orbit conditions. The various root configurations shall be given labels similar to the black-hole case, with the main difference being that, instead of a horizon, we have the tip of the spacetime $r=\beta$. We note that 
\begin{align}
 R(\beta)=-P^2\beta^3(\beta-\alpha)\leq0.
\end{align}
In other words, as long as $P$ is non-zero, the spacetime tip is not accessible to the particle. We modify our sketch of $R(r)$ for the various cases accordingly to obtain Fig.~\ref{fig_top_domain}. The vertical lines, now in green, represent the end-of-spacetime position $r=\beta$.

Looking Eq.~\Eqref{L_sph}, we should keep in mind that this quantity must be positive due to Eq.~\Eqref{domain_QL}. Since now $\beta>\alpha$, this quantity changes sign at
\begin{align}
 \frac{\beta\sqrt{\alpha}}{\sqrt{\alpha}\mp P\sqrt{\beta-\alpha}}\quad\mbox{ and }\quad \frac{3}{2}\alpha.
\end{align}
The root $r=\frac{\beta\sqrt{\alpha}}{\sqrt{\alpha}+P\sqrt{\beta-\alpha}}$ is beyond $r<\beta$, so we need not consider this unphysical location. For the next root, we introduce the notation
\begin{align}
 r_*=\frac{\beta\sqrt{\alpha}}{\sqrt{\alpha} - P\sqrt{\beta-\alpha}}>\beta.
\end{align}
At fixed $\alpha$ and $\beta$, either $r_*$ is larger or smaller than $3\alpha/2$, depending on the value of $P$. We consider each case in turn:

If $0<P<\frac{3\alpha-2\beta}{3\sqrt{\alpha}\sqrt{\beta-\alpha}}$, then $r_*<\frac{3}{2}\alpha$. In this case $K_0$ is positive for $\beta<r<r_*$ and $\frac{3}{2}\alpha<r<\infty$. An example of this situation is shown in Fig.~\ref{fig_top_ELspace_seq1}. The branch $\beta<r<r_*$ corresponds to the middle branch of Fig.~\ref{fig_top_ELspace_seq1}, starting from large positive infinity at $r\rightarrow\beta$, and going to $K=0$ as $r\rightarrow r_*$. For this branch, we find $R''(r)<0$ which correspond to stable circular orbits. The domain $r_*<r<\frac{3}{2}\alpha$ gives negative $K$ so we do not consider it here. Next, increasing $r$ continuously from $\frac{3}{2}\alpha$, we have a stable branch until $r$ reaches the critical point $r_{\mathrm{CCO}}$. This point correspond to the sharp cusp in Fig.~\ref{fig_top_ELspace_seq1}. After this critical point, the remaining branch $r_{\mathrm{CCO}}<r<\infty$ is a branch of stable circular orbits. In summary, for the TS case we have two branches of stable circular orbits, $\beta<r<r_*$ and $r_{\mathrm{CCO}}<r<\infty$.

If $\frac{3\alpha-2\beta}{3\sqrt{\alpha}\sqrt{\beta-\alpha}}<P<P_{\mathrm{crit}}$ for some $P_{\mathrm{crit}}$, then $r_*>\frac{3}{2}\alpha$. Here we have a stable branch from $r_*$, increasing until a critical point, which we shall denote by $r_{\mathrm{CCO}1}$. The point $\brac{K(r_{\mathrm{CCO}1}),E^2_0(r_{\mathrm{CCO1}})}$ is the upper-right sharp cusp of Fig.~\ref{fig_top_ELspace_seq2}. As $r$ increases further, we get the unstable branch which ends at another critical point $r_{\mathrm{CCO2}}$, where its corresponding point is the lower-left cusp in Fig.~\ref{fig_top_ELspace_seq2}. After which we reach a stable branch $r_{\mathrm{CCO2}}<r<\infty$. As $P$ increases toward $P_{\mathrm{crit}}$, the two critical points $r_{\mathrm{CCO1}}$ and $r_{\mathrm{CCO2}}$ approach each other, as can be seen in going from Fig.~\ref{fig_top_ELspace_seq2} to Fig.~\ref{fig_top_ELspace_seq3}.

As $P\rightarrow P_{\mathrm{crit}}$, the two critical points coalesce. For the example of $\alpha=1.5$ and $\beta=2$, the value is about $P_{\mathrm{crit}}\approx0.354$. 
Increasing $P$ further beyond that results in a smooth curve separating domains A2 and A3 (see Fig.~\ref{fig_top_ELspace_seq4}), and this single branch is stable. This is highly reminiscent to the swallow-tail graphs of thermodynamics depicting phase transitions. 

Finally, as $P$ increases further toward $\frac{\alpha}{\beta-\alpha}$, the single circular orbit branch approaches the line $E^2=1+P^2$, thus shrinking the domain A2. If $P$ continues to increase beyond that, $K$ in Eq.~\Eqref{L_sph} becomes negative for any $r$ and no circular orbits can occur.

\subsubsection{\texorpdfstring{$\alpha<\frac{2}{3}\beta$}{alpha<2beta/3} (TS)}

\begin{figure}
 \begin{center}
  \begin{subfigure}[b]{0.49\textwidth}
    \centering
    \includegraphics[scale=1]{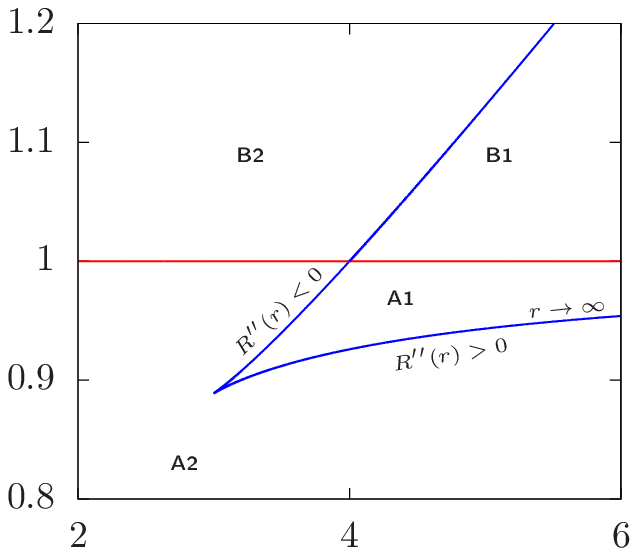}
    \caption{$P=0$.}
    \label{fig_top2_ELspace_seq1}
  \end{subfigure}
  \begin{subfigure}[b]{0.49\textwidth}
    \centering
    \includegraphics[scale=1]{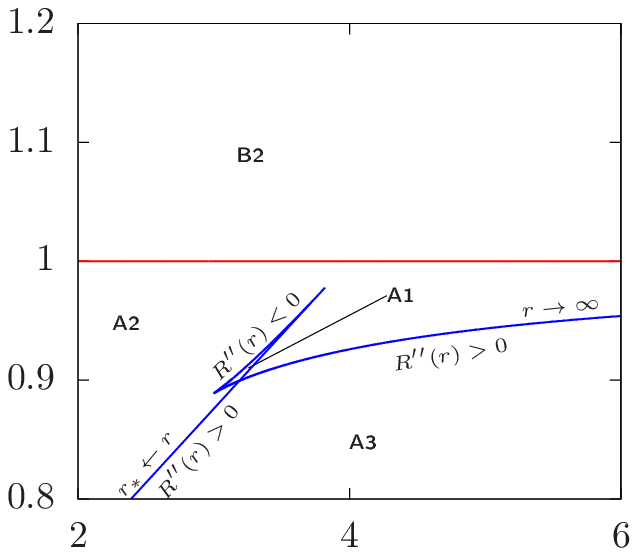}
    \caption{$P=0.002$.}
    \label{fig_top2_ELspace_seq2}
  \end{subfigure}
  \begin{subfigure}[b]{0.49\textwidth}
    \centering
    \includegraphics[scale=1]{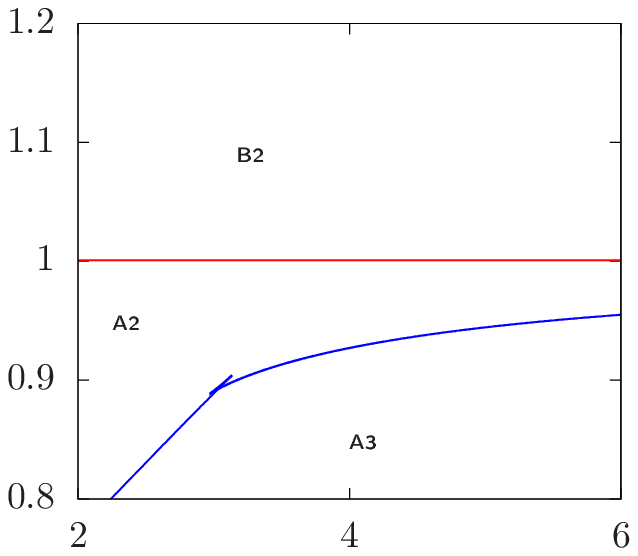}
    \caption{$P=0.05$.}
    \label{fig_top2_ELspace_seq3}
  \end{subfigure}
  \begin{subfigure}[b]{0.49\textwidth}
    \centering
    \includegraphics[scale=1]{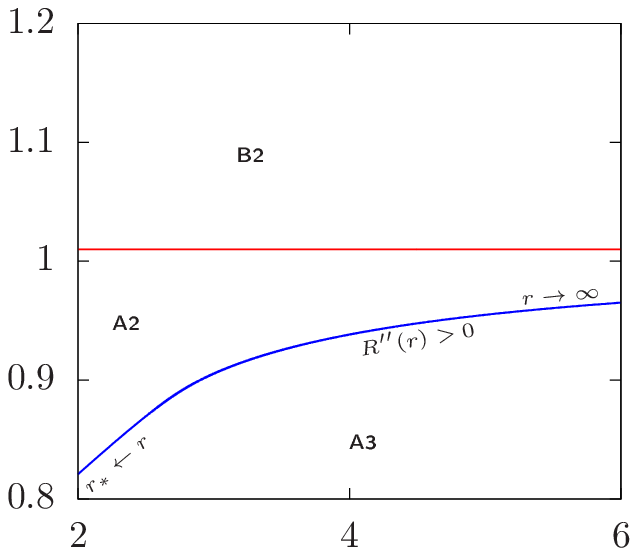}
    \caption{$P=0.1$.}
    \label{fig_top2_ELspace_seq4}
  \end{subfigure}
 \caption{Parameter space for the soliton/topological star with $\alpha=1$, $\beta=2$, and various $P$. The horizontal direction represents $K=L^2+Q$ and the vertical direction represents $E^2$. The horizontal red lines correspond to $E^2=1+P^2$.}
 \label{fig_top2_ELspace_seq}
 \end{center}

\end{figure}

For $\beta>\frac{3}{2}\alpha$, the position $r=\frac{3}{2}\alpha$ is beyond the physical range. Then $K_0$ only changes sign at $r_*>\beta$. In this case, the situation depicted in Fig.~\ref{fig_top_ELspace_seq1} does not exist here. The remaining sequence of structure as $P>0$ is gradually increases is similar to the previous case, and is shown in Fig.~\ref{fig_top2_ELspace_seq}, in that there are two stable branches and one unstable branch. As $P$ reaches $P_{\mathrm{crit}}$, the unstable branch disappears and the two stable branch merges into one, shown in Fig.~\ref{fig_top2_ELspace_seq4}. For the example $\alpha=1$, $\beta=2$, the value of the critical momentum is about $P_{\mathrm{crit}}\approx0.53$.

\section{Examples of orbits and special cases}

\subsection{The \texorpdfstring{Poincar\'{e}}{Poincare} cone} 

In Sec.~\ref{subsec_angular}, it was shown that the trajectory of the orbits lie on a Poincar\'{e} cone whose orientation $\psi$ and opening angle $\chi$ is given by Eq.~\Eqref{cone_psi} and \Eqref{cone_chi}, respectively. In this section, we shall explore some concrete examples and plot the orbits, thus demonstrating the geometrical significance of the cone.

From Eqs.~\Eqref{cone_psi} and \Eqref{cone_chi}, we see that the angles $\psi$ and $\chi$ are determined by $L$, $Q$, and $q$. We first construct some examples of circular orbits whose cones lie parallel to the $x_3$-axis. For this we require $\psi=0$ or $\psi=\pi$. This is achieved when $Q=-q^2$. For constant $r$, the requisite values of $E$ and $K$ are determined by Eq.~\Eqref{E_sph} and \Eqref{L_sph}. Then the appropriate value of $L$ is chosen to satisfy the equation $L^2=K-Q$.

\begin{figure}
 \begin{center}
  \includegraphics{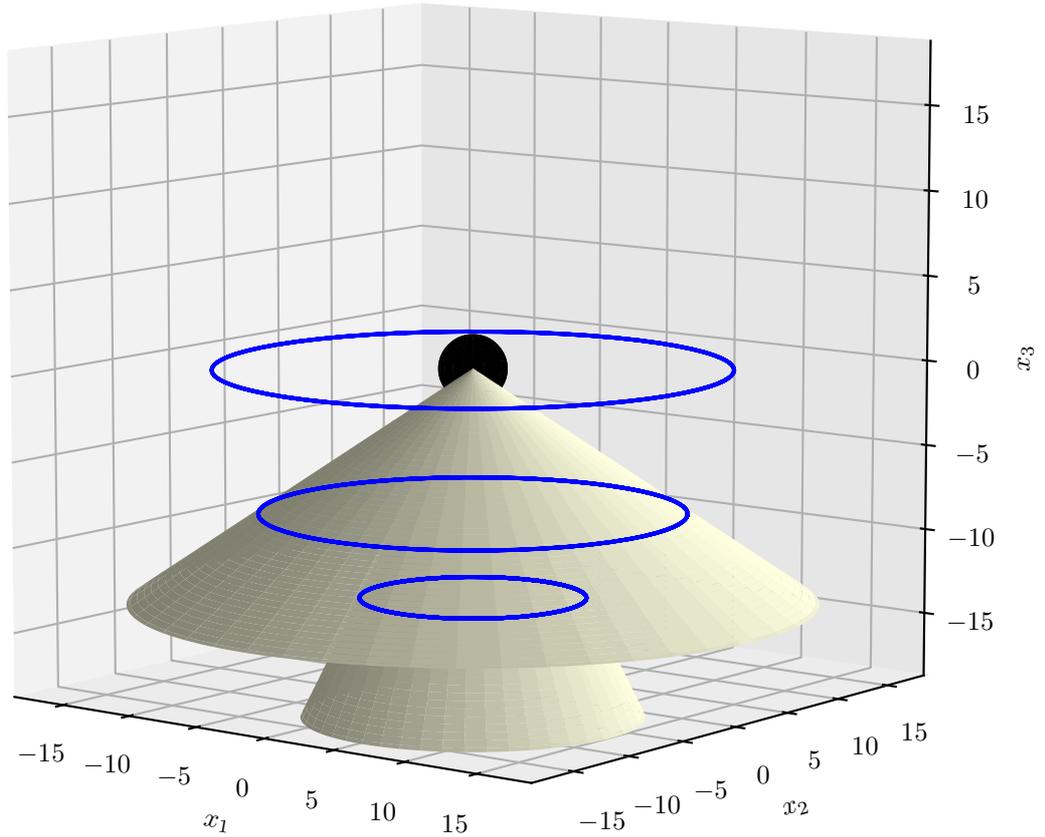}
  \caption{Three spherical orbits of radius $r=15$ and momentum $P=0.2$ around a black hole spacetime of parameters $\alpha=2$, $\beta=1$. From top to bottom, the values of $q$ are $0$, $3$, and $9$. The values of $E$ and $K$ are as calculated from Eq.~\Eqref{E_sph} and \Eqref{L_sph}. In each case, $Q=-q^2$, ensuring the cone is vertical and the orbits are all parallel to the $x_1$-$x_2$ plane. The $q=0$ orbit lies on a `cone' of opening angle $\pi$, which is simply the $x_3=0$ plane. For $q=3$ and $q=9$, their opening angles are $\chi=1.9404\;\mathrm{rad}$ and $\chi=0.90577\;\mathrm{rad}$ respectively, as determined from \Eqref{cone_chi}.}
  \label{fig_ConePlayCircular1}
 \end{center}

\end{figure}

\begin{figure}
 \begin{center}
  \begin{subfigure}[b]{0.49\textwidth}
    \centering
    \includegraphics[width=\textwidth]{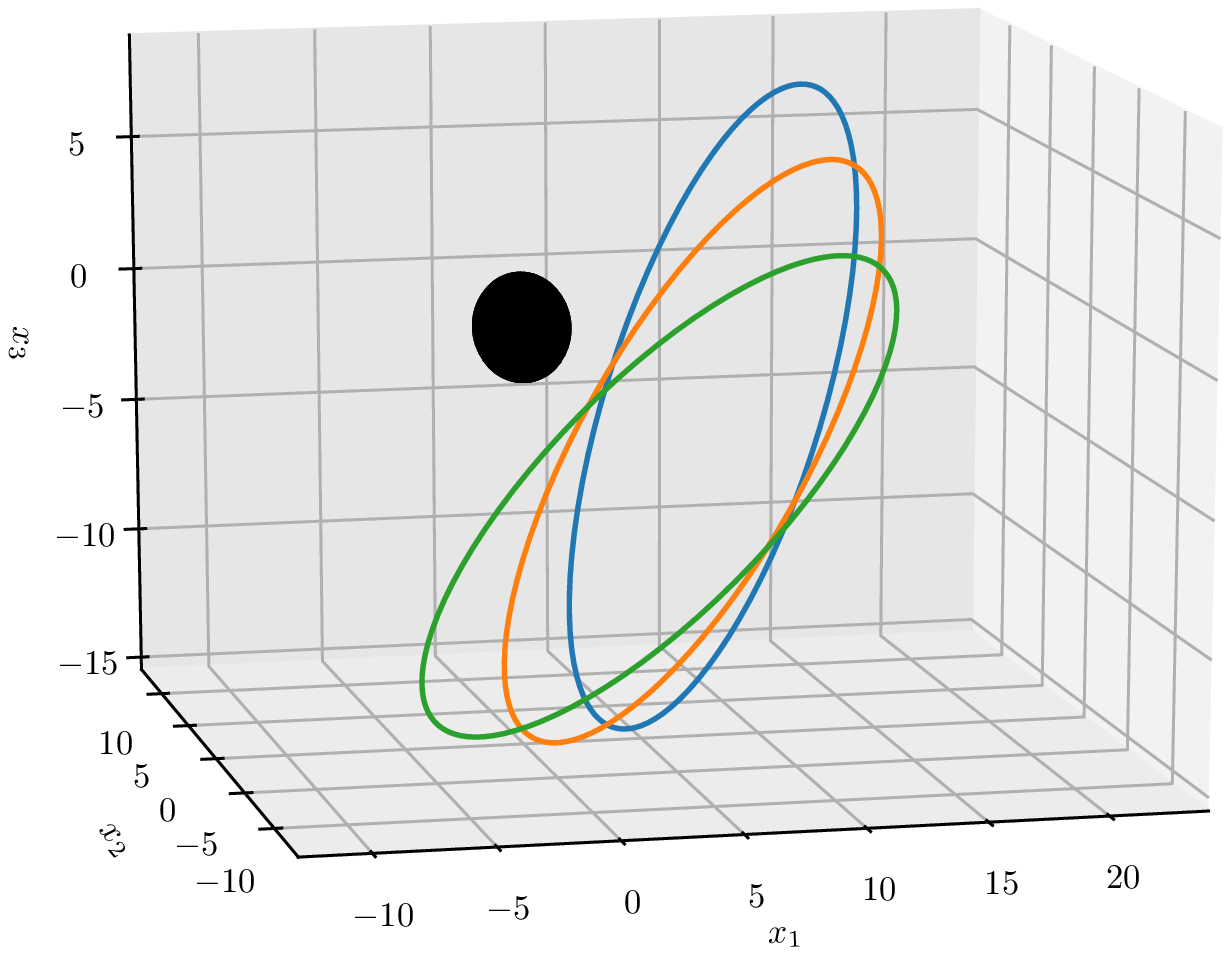}
    \caption{Constant $r=15$.}
    \label{fig_AzConf1}
  \end{subfigure}
  \begin{subfigure}[b]{0.49\textwidth}
    \centering
    \includegraphics[width=\textwidth]{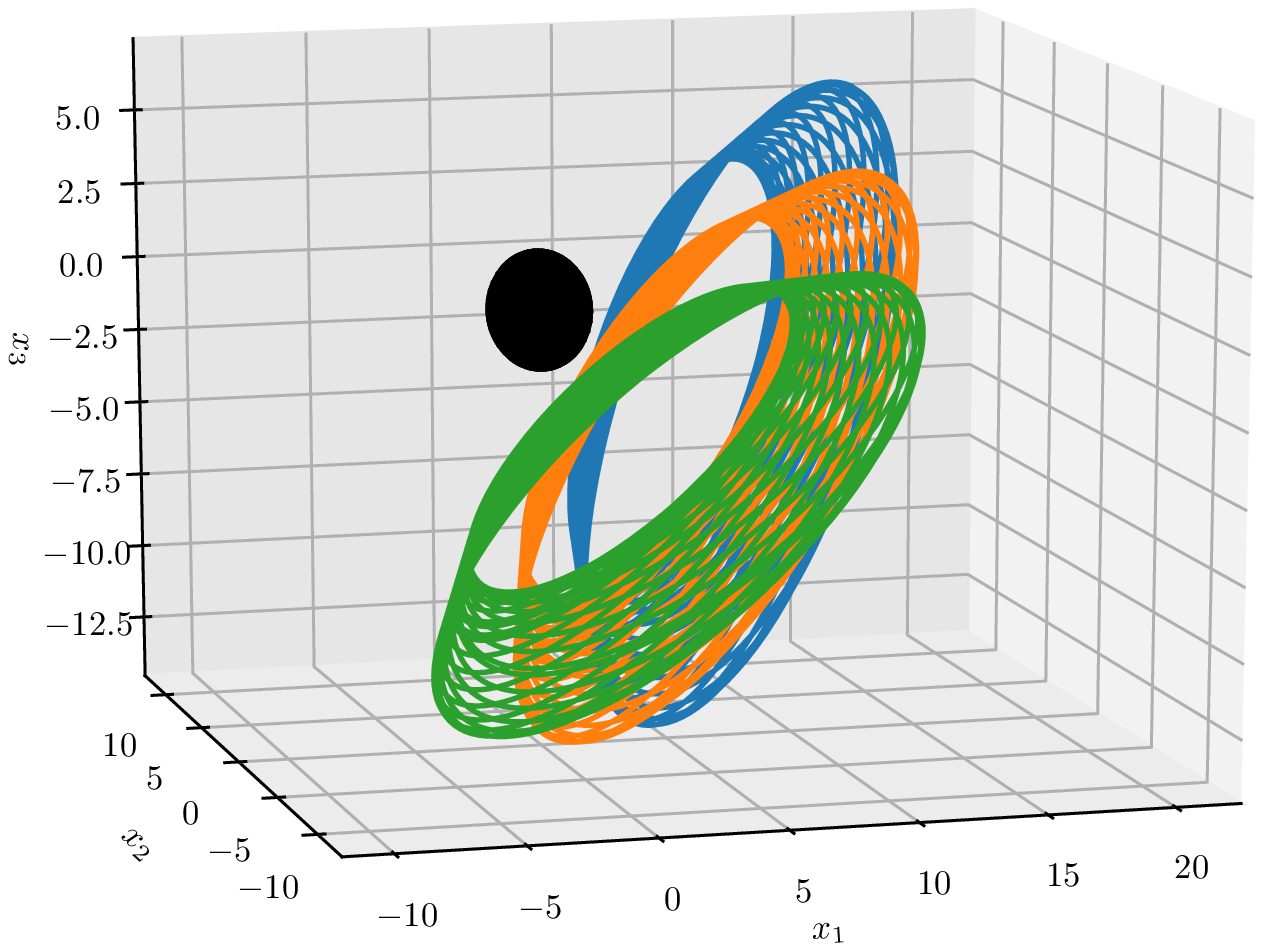}
    \caption{$E=\sqrt{0.985}$, $K=16$.}
    \label{fig_AzConf2}
  \end{subfigure}
  \caption{Orbits around a black hole spacetime of $\alpha=2$ and $\beta=1$ of a particle of charge $q=3$ and momentum $P=0.25$. From highest to lowest, the angular momenta are $L=2$ (blue), $L=3$ (green), and $L=4$ (orange). For Fig.~\ref{fig_AzConf1}, the values of $E$ and $K$ are determined from Eq.~\Eqref{EL_sph}. This time their underlying cones are not plotted to avoid cluttering the figure.}
  \label{fig_AzConf}
 \end{center}

\end{figure}

Starting with $q=0$, we have a `cone' with opening angle $\chi=\pi$, which is simply the flat $x_1$-$x_2$ plane. When $q$ is increased, the opening angles change in accordance to Eq.~\Eqref{cone_chi}. Some examples are shown in Fig.~\ref{fig_ConePlayCircular1}. Next if we wish to consider fixed opening angles, Eq.~\Eqref{cone_chi} tells us that $\chi$ can be fixed if $q$ and $K=Q+L^2$ are fixed. So varying $L$ will now change the angle of the cone with respect to the $x_3$-axis. Some examples of this, with constant and non-constant $r$ are shown in Fig.~\ref{fig_AzConf}.

\subsection{Bound orbits around a TS}

When $\alpha<\beta$ we have seen in the previous section that there exist two distinct domains of bound orbits, which is denoted as Case A1 and sketched in Fig.~\ref{fig_top_domainA1}. Changing the energy and angular momentum may cause two of the roots to coalesce and become complex, leaving a single connected domain of bound orbits, denoted Case A2 and sketched in Fig.~\ref{fig_top_domainA2}.

\subsubsection{`Low energy' bound orbits (\texorpdfstring{$E<\sqrt{1+P^2}$}{E<sqrt{1+P2}})}

Let us first consider Case A1 in further detail. In this case, $R(r)$ has four real roots, which we denote to have the following order:
\begin{align}
 r_1\leq r_2\leq r_-\leq r_+.
\end{align}
In Case A1, we have bound orbits where the particle can exist either in the interval $[r_1,r_2]$ or $[r_-,r_+]$. If we place our particle with initial conditions $r(0)=r_1$ and $\frac{\dif r}{\dif\lambda}(0)=0$, the trajectory of the particle is given by the analytical solution 
\begin{align}
 r(\lambda)&=\frac{(r_+-r_2)r_1+(r_2-r_1)r_+\mathrm{sn}\brac{\eta\lambda,p}^2}{r_+-r_2+(r_2-r_1)\mathrm{sn}\brac{\eta\lambda,p}^2},
\end{align}
where $\mathrm{sn}\brac{\psi,k}$ is the Jacobi elliptic function of the first kind, and
\begin{align}
 \eta=\half\sqrt{(1+P^2-E^2)(r_+-r_2)(r_--r_1)},\quad p=\sqrt{\frac{(r_+-r_-)(r_2-r_1)}{(r_+-r_2)(r_--r_1)}}. \label{eta_p_def}
\end{align}

\begin{figure}
 \begin{center}
  \includegraphics[scale=0.8]{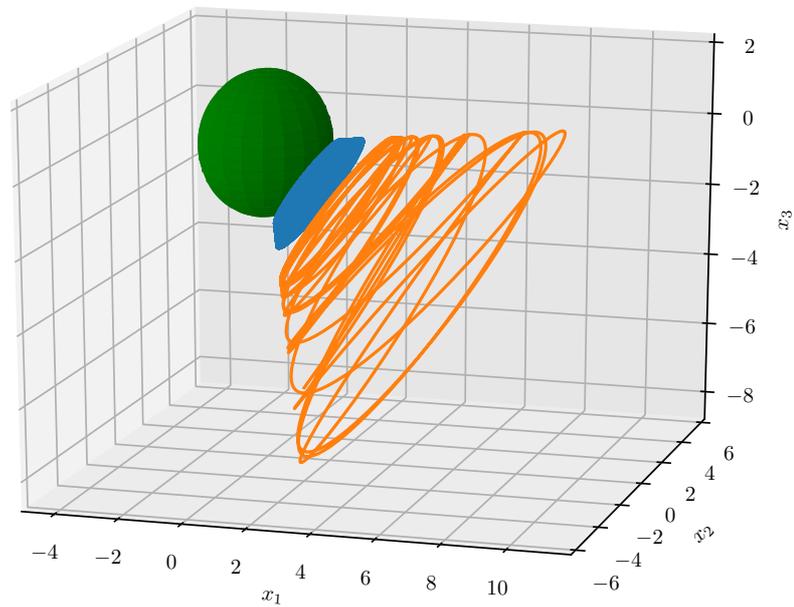}
  \caption{Orbits around a topological star spacetime with $\alpha=1.5$, $\beta=2$, for a particle with $q=3$, $P=0.25$, $E=\sqrt{0.97}$, $K=6.8$, $Q=1$, and $L=\sqrt{K-Q}$. The particle in the inner domain $r_1<r<r_2$ is depicted with a blue curve, where $r_1=2.24334$ and $r_2=2.87916$. The particle with the outer domain $r_-\leq r\leq r_+$ is shown by the blue curve, where $r_-=3.82194$ and $r_+=8.89339$. The solid sphere represents the surface $r=\beta=2$.}
  \label{fig_orbit_plotting_top_A1}
 \end{center}

\end{figure}

To describe bound orbits in the outer domain, let us choose the initial conditions $r(0)=r_-$ and $\frac{\dif r}{\dif\lambda}(0)=0$. Then its trajectory will be described by the solution 
\begin{align}
 r(\lambda)&=\frac{(r_+-r_2)r_--(r_+-r_-)r_2\mathrm{sn}\brac{\eta\lambda,p}^2}{r_+-r_2+(r_--r_+)\mathrm{sn}\brac{\eta\lambda,p}^2}.
\end{align}
An example of orbits in these two domains are shown in Fig.~\ref{fig_orbit_plotting_top_A1}.

\begin{figure}
 \begin{center}
  \includegraphics[scale=0.8]{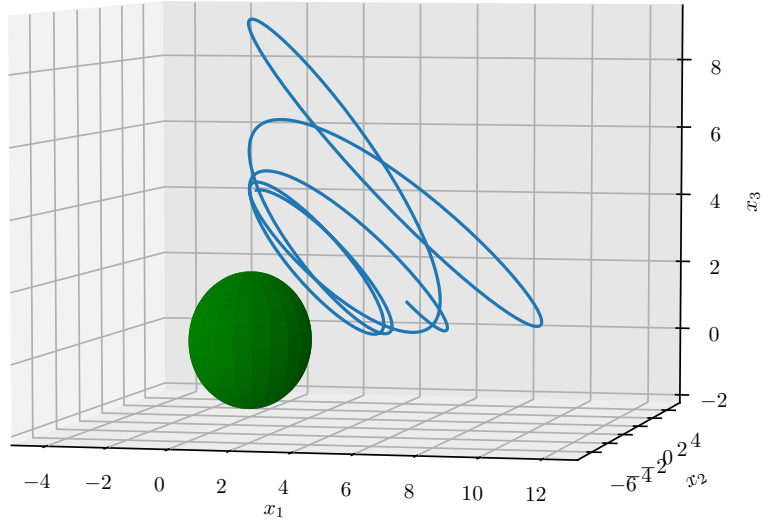}
  \caption{Orbits around a topological star spacetime with $\alpha=1.5$, $\beta=2$, for a particle with $q=3$, $P=0.25$, $E=\sqrt{0.98}$, $K=7.2$, $Q=-1$, and $L=-\sqrt{K-Q}$. The particle moves in the domain $r_1<r<r_2$ where $r_1=4.3744$ and $r_2=10.7069$. The other two roots of $R(r)$ are complex and are given by $m\pm\im n$, where $m=2.3609$ and $n=0.1279$. The solid sphere represents the surface $r=\beta=2$.}
  \label{fig_orbit_plotting_top_A2}
 \end{center}

\end{figure}

In case A2, the function $R(r)$ has two real roots, which we denote by $r_1\leq r_2$. We further denote the other pair of complex conjugate roots as $m\pm\im n$. Then the function $R(r)$ is written as 
\begin{align}
 R(r)=(1+P^2-E^2)(r_2-r)(r-r_1)\sbrac{(r-m)^2+n^2}.
\end{align}
The particle can exist in the domain $r_1\leq r\leq r_2$ where $R(r)\geq0$. Choosing initial conditions $r(0)=r_1$ and the upper sign for Eq.~\Eqref{dr}, the analytical solution is 
\begin{align}
 r(\lambda)&=\frac{Br_2+Ar_1\cot\left\{\half\arcsin\sbrac{\mathrm{sn}\brac{\delta\lambda,\rho}} \right\}}{B+A\cot\left\{\half\arcsin\sbrac{\mathrm{sn}\brac{\delta\lambda,\rho}} \right\}},
\end{align}
where 
\begin{align}
 A&=(m-r_2)^2+n^2,\quad B=(m-r_1)^2+n^2,\nonumber\\
 \delta&=\sqrt{AB\brac{1+P^2-E^2}},\quad\rho=\half\sqrt{\frac{-(A-B)^2+(r_2-r_1)^2}{AB}}.
\end{align}
An example of such an orbit is shown in Fig.~\ref{fig_orbit_plotting_top_A2}

\subsubsection{`High energy' bound orbits (\texorpdfstring{$E>\sqrt{1+P^2}$}{E>sqrt{1+P2}})}

\begin{figure}[h]
 \begin{center}
  \includegraphics[scale=0.7]{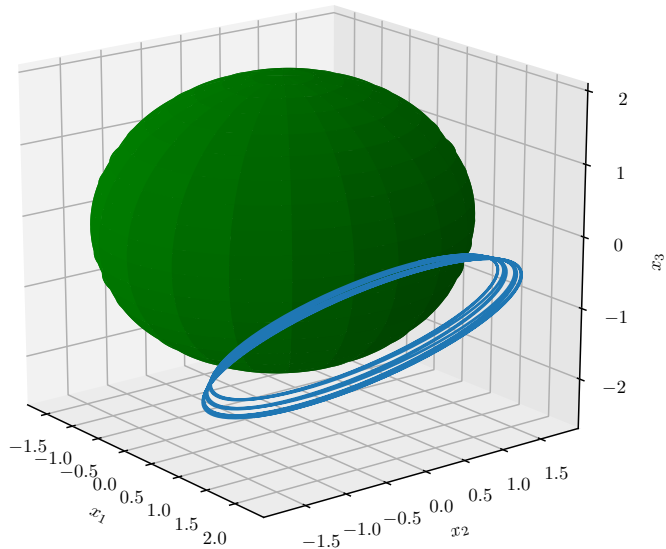}
  \caption{Orbits around a topological star spacetime with $\alpha=1.5$, $\beta=2$, for a particle with $q=3$, $P=0.25$, $E=\sqrt{1.066}$, $K=8.44$, $Q=1$, and $L=\sqrt{K-Q}$. The particle moves in the domain $r_2<r<r_-$ where $r_2=2.3576$ and $r_-=2.6231$. The solid sphere represents the surface $r=\beta=2$.}
  \label{fig_Orbit-B1}
 \end{center}

\end{figure}

For particles of `high' energy $E>\sqrt{1+P^2}$, bound orbits may still exist for sufficiently small $P$ where case B1 exists. In this case, the function $R(r)$ has four real roots $r_1\leq r_2\leq r_-\leq r_+$ and is non-negative in the domain $r_2\leq r\leq r_-$. The leading coefficient of $R(r)$ is positive. So the function can be written as 
\begin{align}
 R(r)=\brac{E^2-1-P^2}(r-r_1)(r-r_2)(r_--r)(r_+-r).
\end{align}
Choosing initial conditions $r(0)=r_2$ and the upper sign in Eq.~\Eqref{dr}, the analytical solution is 
\begin{align}
 r(\lambda)&=\frac{(r_--r_1)r_2-(r_--r_2)r_1\mathrm{sn}\brac{\zeta\lambda,b}^2}{r_--r_1-(r_--r_2)\mathrm{sn}\brac{\zeta\lambda,b}^2},
\end{align}
where
\begin{align}
 \zeta=\half\sqrt{(E^2-P^2-1)(r_+-r_2)(r_--r_1)} =\im\eta,\quad b=\sqrt{\frac{(r_--r_2)(r_+-r_1)}{(r_+-r_2)(r_--r_1))}}.
\end{align}
An example of such an orbit is shown in Fig.~\ref{fig_Orbit-B1}.

\section{Conclusion} \label{sec_conclusion}

In this paper we have derived and analysed the equations of motion for a charged particle in the magnetic black hole/topological star solution. The angular motion was found to be similar to analogous systems involving interacting electric and magnetic monopoles. In particular, the electric charge moves along the surface of a Poincar\'{e} cone. The angle and orientation of the cone depends on the charge and the angular momentum, as is well-known since Poincar\'{e}'s original non-relativistic analysis. 

On the other hand, the radial motion shows a richer structure of possibilities in the topological star case. In particular, up to two distinct domains of bound orbits may exist for fixed energy and total angular momentum. On the $E^2$-$K$ space, the points representing circular orbits exhibit a swallow-tail structure where each branch correspond to stable/unstable circular orbits. When the $w$-momentum is varied, the swallow-tail kink disappears, leaving just a single stable branch. This is highly reminiscent of phase transitions behaviour in thermodynamics, where swallow-tail structures appear in the graphs of intrinsic parameters. Investigating this similarity and the possibility of carrying over thermodynamic concepts into particle mechanics of this spacetime might be a candidate of future study.


\bibliographystyle{topstar}

\bibliography{topstar}

\end{document}